\documentclass{article}
\usepackage{format}
\usepackage{macros}
\usepackage{paper}

\title{On the correctness of Egalitarian Paxos}

\author{Pierre Sutra}

\affil{
    T\'el\'ecom SudParis\\
    9, rue Charles Fourier\\
    91000 \'Evry, France
}

\date{}

\begin{document}

\maketitle

\begin{abstract}
  This paper identifies a problem in both the \tlaplus specification and the implementation of the Egalitarian Paxos protocol.
  It is related to how replicas switch from one ballot to another when computing the dependencies of a command.
  The problem may lead replicas to diverge and break the linearizability of the replicated service.
\end{abstract}


\section{Introduction}
\labsection{introduction}

State-machine replication (SMR) is a fundamental technique to build dependable services.
With SMR, a service is replicated across a set of distributed processes.
Replicas apply to their local copies the commands that access the replicated service following a total order.
This order is implemented using a repeated agreement protocol, or \emph{consensus}, on the next command to execute.

Some recent works \cite{gbroadcast,gpaxos} observe that it is not necessary to build a total order over the commands submitted to the service.
To maintain service consistency and provide the illusion of having no data replication, ordering non-commuting commands suffices.
Egalitarian Paxos (\epaxos) \cite{epaxos} is a novel protocol built upon this insight.
To construct the partial order, \epaxos replicas agree on the dependencies of each command submitted to the service.
A replica executes commands according to the graph formed by these dependencies, that is, if $c$ depends on $d$, then $c$ is executed after $d$.

To agree on the dependencies of a command, \epaxos follows a common pattern:
replicas successively join asynchronous ballots, and during a ballot, they try to make a decision.

Surprisingly, the \tlaplus specification and the Golang implementation of \epaxos use a single variable at each replica to track progress across ballots.
This paper shows that this is not sufficient.
We exhibit an admissible execution in which replicas disagree on the dependencies of a command, breaking consistency.

\paragraph{Outline}
\refsection{consensus} recalls the traditional schema of repeated asynchronous ballots to reach consensus.
\refsection{epaxos} gives an overview of the \epaxos algorithm.
\refsection{counterexample} depicts our counter-example.
\refsection{conclusion} closes this paper.
An appendix follows that contains the \tlaplus specification of the counter-example.
The specification is also available online \cite{on-epaxos-correctness}.

\section{Solving consensus}
\labsection{consensus}

The classical approach to solve consensus is to execute a sequence of asynchronous rounds, or \emph{ballots}.
Each ballot is identified with a natural, its \emph{ballot number}.
As usual, we refer to a ballot with its ballot number and assume that processes start the agreement at ballot $0$.

During a ballot, a \emph{quorum} of processes attempt to agree on some proposed value.
To this end, each ballot is split into three distinct phases.
Before it participates in a ballot, a process first \emph{joins} it (the \emph{prepare} phase).
Then, it may \emph{vote} for some proposed value (\emph{accept} phase).
A value is \emph{chosen} at a ballot when a quorum of processes voted for it.
A process \emph{decides} some value once it knows that this value was chosen at some ballot (\emph{learn} phase).

The size of a quorum depends on the time taken by the protocol to reach a decision in a ballot.
Consider a system of $n=2f+1$ processes of which at most $f$ may fail-stop.
In Paxos \cite{paxos}, any majority of the processes (that is, at least $f+1$ of them) is a quorum at every ballot.
Differently, Fast Paxos \cite{fastpaxos} distinguishes fast and classic ballots.
Whereas a majority of processes is a quorum for a classic ballot, a quorum of a fast ballot contains at least $f + \lfloor\frac{f+1}{2}\rfloor + 1$ processes.

Reaching consensus requires to choose a unique value among the proposals.
To this end, the following invariant is maintained across ballots during an execution:
\begin{itemize}
\item[] (INV) if a value $v$ is chosen at some ballot $b$, then for every ballot $b' \geq b$, if $u$ is chosen at $b'$, $u=v$ holds.
\end{itemize}
To build this invariant, consensus algorithms commonly rely on the following assumptions.
\begin{itemize}
\item[(A1)]%
  A process can join a ballot $b$ only if it did not join some ballot $b'>b$ previously.    
\item[(A2)]%
  A process may only cast a vote for the last ballot it has joined.
\item[(A3)]%
  The vote of a process at a ballot is irrevocable.
  This means that if a process votes for some value $u$ at ballot $b$, it cannot later vote at ballot $b$ for some other value $v \neq u$.
\item[(A4)]%
  The votes of any two processes at a ballot are identical, that is, if $p$ and $q$ vote at ballot $b$ for respectively $u$ and $v$, then $u = v$.
\item[(A5)]%
  Consider two quorums $Q$ and $Q'$ defined respectively at ballot $b$ and $b'$.
  Then, the intersection of $Q$ and $Q'$ is non-empty.
\end{itemize}

To maintain invariant (INV), the consensus protocol tracks the values chosen at prior ballots when it progresses to a new ballot.
In detail, a process $p$ that aims at advancing to ballot $b$, first seals prior ballots.
To this end, for every ballot $b'<b$, and every quorum $Q$ at $b'$, $p$ asks a process $q \in Q$ to join $b$.
By invariants (A1) and (A2), sealing prevents new value to be chosen at a lower ballot than $b$.

Once a quorum $Q$ of processes have joined ballot $b$, $p$ computes the values chosen prior to $b$.
Assume that some value $v$ is chosen at ballot $b-1$ by a quorum $Q'$.
By invariants (A3-A5), $v$ is unique, and some process $q \in Q \inter Q'$ voted for $v$ at ballot $b-1$.
Hence, process $p$ discovers $v$ when it inquiries the processes in $Q$ at the time they join ballot $b$.

Now if $v$ does not exist, by induction, $p$ must propose what was voted at ballot $b-2$, etc.
In other words, $p$ proposes at ballot $b$ the value for which some process in $Q$ voted at the largest ballot prior to $b$.
If no such value exists, $p$ is free to pick any value as its proposal.
If (INV) is true up to ballot $b-1$, then the invariant is maintained at ballot $b$.

The proposal of $p$ is a so-called \emph{safe} value.
It maintains invariant (INV) at ballot $b$ by over-approximating what was chosen at prior ballots.
In the literature \cite{gpaxos}, the construction of a safe value makes use of three variables at each process.
The first one, denoted hereafter $\bal$, is the last ballot joined so far by a process.
The second variable $\vbal$ is the last ballot at which the process voted for some value.
Variable $\vval$ contains the value that was voted at ballot $\vbal$.

The \tlaplus specification of \epaxos and its implementation rely on not two ballot variables, as mentioned above, but a single one to reach an agreement on the dependencies of a command.
In the remainder of this paper, we explain that this mechanism is flawed.
To this end, we give an overview of the \epaxos protocol, then we exhibit an admissible run that breaks safety.

\section{Overview of \epaxos}
\labsection{epaxos}

Egalitarian Paxos (\epaxos) is a recent protocol to implement state-machine replication and construct fault-tolerant distributed services.
\epaxos is leaderless and it orders commands in a decentralized way without relying on a distinguished process.
As in \cite{gbroadcast, gpaxos}, \epaxos exploits the commutativity between commands submitted to the replicated service to improve performance.
In the most favorable case, that is when there is no concurrent non-commuting command, the protocol decides (aka., commit) the next command to execute after one round-trip to the closest fast quorum.

With more details, processes running the \epaxos protocol agree on a directed graph, or \emph{execution graph}.
Commands are executed on the local copy of the service following some linearization of this graph.%
\footnote{
  Cycles in the execution graph are broken deterministically.
}
To build the execution graph, for each command $c$ submitted to the system, \epaxos constructs a set of dependencies $\dep(c)$.
The dependencies of $c$ precede it in the execution graph.
Command $c$ is \emph{committed} at some process $p$ once $\dep(c)$ is known at $p$.
A command is executed once the transitive closure of its dependencies is committed.

The \epaxos protocol satisfies two core invariants detailed below.
These invariants ensure that the protocol maintains the consistency of the replicated service, namely its linearizability \cite{linearizability}.
\begin{itemize}
\item[(E1)] Any two processes agree on the dependencies of a command.
\item[(E2)] For any two non-commuting committed commands $c$ and $d$, $c \in \dep(d)$ or the converse holds.
\end{itemize}

To agree on the dependencies of a command, \epaxos employs a variation of Fast Paxos.
The first (so-called) \emph{pre-accept} phase of the agreement is not coordinated.
More precisely, to propose some command $c$, a process $p$ propagates $c$ to a quorum.
Upon receiving command $c$, a process sends all the commands conflicting with $c$ (that it is aware of) back to $p$.
If all the processes return the same dependencies, a spontaneous agreement occurs.
Otherwise, a classical accept phase takes place.

Let us notice that when ballot $b$ is not coordinated, processes may return different values for $\dep(c)$, and invariant A4 does not hold anymore.
When $b$ is fast, (A4) is replaced with the following Fast Paxos invariant \cite{fastpaxos}:
\begin{itemize}
\item[(A4')]%
  For any two quorums $Q$ and $Q'$ at ballot $b$, and any quorum $Q''$ at some ballot $b'$, $Q \inter Q' \inter Q'' \neq \emptySet$.
\end{itemize}
Invariant (A4') ensures that a process starting ballot $b'$ (a \emph{recovery} in \epaxos parlance) observes at most a single chosen value at ballot $b$, the one with the most votes.

Differently from Fast Paxos, only ballot $0$ is fast in \epaxos and allows a spontaneous agreement to form.%
\footnote{
  During a recovery, the TryPreAccept messages serve solely to ensure invariant (E2).
} 
In addition, when a process proposes a command, it includes its local computation of $\dep(c)$ in its proposal.
This allows to reduce the size of a fast quorum at ballot $0$ to $f+\lfloor\frac{f+1}{2}\rfloor$ processes.

\section{Breaking safety}
\labsection{counterexample}

\begin{figure*}[t]
  \centering
  \fontsize{8}{11}\selectfont

  \begin{tabular}{lr}
    \subfigure[Key states]{
      \begin{tabular}{c|c|c}
        Step & Process & $(\status,\bal,\dep)$ for $c_2$ in $\langle p_1, 1 \rangle$ \\
        \hline
        \hline
        7 & $p_3$ & $\flag{accepted}, 1, \{c_1\}$ \\
        \hline
        15 & $p_3$ & $\flag{accepted}, 1, \{c_1\}$ \\
        & $p_2$ & $\flag{accepted}, 2, \{\}$ \\
        & $p_1$ & $\flag{accepted}, 2, \{\}$ \\
        \hline
        17 & $p_3$ & $\flag{accepted}, 3, \{c_1\}$ \\
        & $p_2$ & $\flag{accepted}, 2, \{\}$ \\
        & $p_1$ & $\flag{accepted}, 2, \{\}$ \\
        \hline
        24 & $p_3$ & $\flag{accepted}, 4, \{c_1\}$ \\
        & $p_2$ & $\flag{committed}, 2, \{\}$ \\
        & $p_1$ & $\flag{committed}, 4, \{c_1\}$
        \bigskip
      \end{tabular}
    }
    \vspace{3em}
    \subfigure[Message labels]{
      \begin{tabular}{c|c}
        Color & Type \\
        \hline
        \hline
        \color{Black}{$\rightarrow$} & \flag{pre-accept} \\
        \color{Gray}{$\rightarrow$} & \flag{pre-accept-reply} \\
        \color{Red}{$\rightarrow$} & \flag{prepare} \\
        \color{Orange}{$\rightarrow$} & \flag{prepare-reply} \\
        \color{OliveGreen}{$\rightarrow$} & \flag{accept} \\
        \color{Green}{$\rightarrow$} & \flag{accept-reply}
        \bigskip
      \end{tabular}
    }
  \end{tabular}

  \subfigure[Execution timeline]{

    \begin{tabular}{c}

      \begin{tikzpicture}
        \path[draw,color=gray] (0.5,1) edge (11.5,1);
        \path[draw,color=gray] (0.5,1.8) edge (11.5,1.8);
        \path[draw,color=gray] (0.5,2.6) edge (11.5,2.6);

        \path[draw,dashed,color=gray] (11.5,1) edge (12,1);
        \path[draw,dashed,color=gray] (11.5,1.8) edge (12,1.8);
        \path[draw,dashed,color=gray] (11.5,2.6) edge (12,2.6);
        
        \node at (0.2,1) {$p_1$};
        \node at (0.2,1.8) {$p_2$};
        \node at (0.2,2.6) {$p_3$};

        \node[above] at (0.9,2.6) {1};
        \path[preAccept,out=45, in=135] (1,2.6) edge (1.5,2.6);
        
        \node[below] at (1.25,1) {2};
        \path[preAccept] (1.25,1) edge (2,2.6);

        \node[above] at (2.4,2.6) {3};    
        \path[preAcceptOk] (2.4,2.6) edge (3.4,1);

        \node[above] at (3.5,2.6) {4};
        \path[prepare, out=45, in=135] (3.6,2.6) edge (4.3,2.6);
        \path[prepare] (3.6,2.6) edge (4.3,1.8);

        \node[below] at (4.4,1.8) {5};
        \path[prepareReply] (4.5,1.8) edge (5.5,2.6);

        \node[above] at (4.5,2.6) {6};
        \path[prepareReply, out=45, in=135] (4.6,2.6) edge (5.3,2.6);
        
        \node[above] at (5.6,2.6) {\textbf{7}};
        \path[compute] (5.6,2.65) edge (5.6,2.55);

        \node[above] at (5.9,1.8) {8};
        \path[prepare, out=45, in=135] (6,1.8) edge (6.7,1.8);
        \path[prepare] (6,1.8) edge (6.7,1);

        \node[below] at (6.8,1) {9};
        \path[prepareReply] (6.8,1) edge (8,1.8);
        
        \node[above] at (6.9,1.8) {10};
        \path[prepareReply, out=45, in=135] (7,1.8) edge (7.7,1.8);       

        \node[above] at (8.2,1.8) {11};
        \path[preAccept, out=45, in=135] (8.3,1.8) edge (8.8,1.8);
        \path[preAccept] (8.3,1.8) edge (9.3,1);
        
        \node[below] at (9.5,1) {12};
        \path[preAcceptOk] (9.5,1) edge (10,1.8);

        \node[above] at (10.2,1.8) {13};
        \path[accept] (10.2,1.8) edge (10.7,1);    
        
        \node[below] at (10.9,1) {14};
        \path[acceptReply] (10.9,1) edge (11.4,1.8);    

        \pgfresetboundingbox
        \clip[use as bounding box] (0,0) rectangle (13,2.6);
      \end{tikzpicture}

      \\
      
      \begin{tikzpicture}
        \path[draw,dashed,color=gray] (0.5,1) edge (1.5,1);
        \path[draw,dashed,color=gray] (0.5,1.8) edge (1.5,1.8);
        \path[draw,dashed,color=gray] (0.5,2.6) edge (1.5,2.6);

        \path[->,color=gray] (1.5,1) edge (12,1);
        \path[->,color=gray] (1.5,1.8) edge (12,1.8);
        \path[->,color=gray] (1.5,2.6) edge (12,2.6);

        \node[below] at (2.5,1) {\textbf{15}};
        \path[prepare, out=45, in=135] (2.5,1) edge (3.2,1);
        \path[prepare] (2.5,1) edge (3.5,2.6);

        \node[above] at (3.7,2.6) {16};
        \path[prepareReply] (3.7,2.6) edge (4.7,1);
        
        \node[below] at (5,1) {\textbf{17}};
        \path[prepare, out=45, in=135] (5,1) edge (5.6,1);
        \path[prepare] (5,1) edge (6,2.6);
        
        \node[above] at (6.2,2.6) {18};
        \path[prepareReply] (6.2,2.6) edge (7.2,1);
        
        \node[below] at (6.2,1) {19};
        \path[prepareReply, out=45, in=135] (6.2,1) edge (6.9,1);    

        \node[below] at (7.4,1) {20};
        \path[prepareReply, out=45, in=135] (7.4,1) edge (8.1,1);    
        
        \node[below] at (8.3,1) {21};
        \path[accept] (8.3,1) edge (9.3,2.6);
        
        \node[above] at (9.5,2.6) {22};
        \path[acceptReply] (9.5,2.6) edge (10.5,1);
        
        \node[above] at (11,1.8) {23};
        \path[compute] (11,1.75) edge (11,1.85);
        
        \node[above] at (11.5,1) {\textbf{24}};
        \path[compute] (11.5,.95) edge (11.5,1.05);
        
        \pgfresetboundingbox
        \clip[use as bounding box] (0,0) rectangle (13,3);
      \end{tikzpicture}
      
    \end{tabular}
  }
  
  \caption{
    Breaking safety in \epaxos
    \labfigure{counterexample}
  }
\end{figure*}

As detailed in \refsection{consensus}, consensus algorithms commonly employ two ballot variables at a process.
These variables track the last joined ballot ($\bal$) and the last ballot at which a value was voted ($\vbal$).

However, both the \tlaplus specification of \epaxos \cite[pages 109--123]{epaxos-phd} and its implementation \cite{epaxos-implementation} use only $\bal$.%
\footnote{
  The specification and the implementation of \epaxos call \emph{ballot} this unique variable.
}
When a process receives a message to join a ballot $b > \bal$, it sends back $\bal$ and $\vval$, then updates $\bal$ to $b$.
This is surprising as the pseudo-code \cite[Figure 3]{epaxos} seems to correctly identify the pattern, mentioning to send back ``the most recent ballot number accepted'' in the consensus instance.

Avoiding the use of a second ballot variable is not possible.
\reffigure{counterexample} illustrates how to break safety by executing a well-chosen sequence of steps.

In this figure, the system consists of three processes $\{p_1,p_2,p_3\}$.
Processes $p_3$ and $p_1$ propose respectively to compute the dependencies of commands $c_1$ and $c_2$.
This computation takes place respectively in the consensus instances $\langle p_3, 1 \rangle$ and $\langle p_1, 1 \rangle$.
At the end of the execution, the value of $\dep(c_2)$ committed at process $p_2$ is $\emptySet$, while it equals $\{c_1\}$ at processes $p_1$.

In appendix, we provide the \tlaplus code of the execution presented at \reffigure{counterexample}.
This counter-example is located in a set of states which is too large to explore in a reasonable amount of time with the \tlaplus model checker.
As a consequence, the specification in appendix directly injects the admissible execution with the help of a history variable.
The model checker can be then used to validate that the counter-example is actually feasible.
The history variable introduced in appendix is a counter.
Its value coincides with the numbering of the steps in \reffigure{counterexample}.

The disagreement depicted in \reffigure{counterexample} is obtained by executing consecutive recovery phases.
It is based on the following observation:
if $\vbal$ is not used, a process that accepted a value $u$ at a ballot $b$, then later joins $b'>b$ is in a state identical to having accepted $u$ at ballot $b$'.

In detail, the execution at \reffigure{counterexample} consists of the following steps:
\begin{enumerate}
\item[(1-2)] Process $p_3$ proposes command $c_1$, while $p_1$ proposes command $c_2$.
\item[(3)] $p_3$ returns a pre-accept to $p_1$ message with $dep(c_2)=\{c_1\}$.
\item[(4-7)] $p_3$ partially recovers $\langle p_1, 1 \rangle$ by contacting the quorum $\{p_2,p_3\}$.
  This leads to the fact that $dep(c_2)=\{c_1\}$ is accepted at ballot $2$ (step 7).
\item[(8-14)] $p_2$ executes a full recovery of $c_2$ using $\{p_1,p_2\}$.
  Before step 15, $dep(c_2)=\{\}$ is thus accepted at process $p_2$.
\item[(15-16)] $p_1$ executes a partial recovery with quorum $\{p_1,p_3\}$.
  The state of instance $\langle p_1, 1 \rangle$ at process $p_3$ is now $(\flag{accepted}, 3, \{c_1\})$.
\item[(17-22)] $p_1$ executes a full recovery using the same quorum.
  As $p_3$ holds the highest ballot, its value for $\dep(c_2)$ is accepted at ballot $4$.
\item[(23-24)] Processes $p_1$ and $p_2$ commit respectively $\{c_1\}$ and $\emptySet$ for $\dep(c_2)$.
\end{enumerate}

\section{Conclusion}
\labsection{conclusion}

Egalitarian Paxos (\epaxos) is a recent protocol to implement state-machine replication and construct fault-tolerant distributed services.
In the common case, \epaxos delivers a command after one round-trip to the closest fast quorum.
Contrary to prior works, such as Generalized Paxos \cite{gpaxos}, a leader does not need to solve conflicts between non-commuting commands.
These two properties make the protocol particularly appealing for geo-distributed systems.

The repeated consensus procedure of \epaxos must rely on two ballot variables to track the progress at each replica.
This paper shows that if this is not the case, as in the \tlaplus specification and the Golang implementation, breaking safety is possible.

\section*{Acknowledgments}
This research is partly supported by the RainbowFS project of Agence Nationale de la Recherche, France, number ANR-16-CE25-0013-01a, and the European Union's Horizon 2020 research and innovation programme under grant agreement No 825184 (CloudButton).
The author thanks Vitor Enes for his remarks on an initial version of this work, and Alexey Gotsman for fruitful discussions on the Egalitarian Paxos protocol.

\bibliographystyle{elsarticle-num}
\bibliography{bib}

\clearpage
\appendix

\section{The counter example}
\labappendix{counterexample}

\subsection{\tlaplus specification}
\labappendix{counterexample:tla}

\tlatex
\setboolean{shading}{true}
\@x{}\moduleLeftDash\@xx{ {\MODULE} CounterExample}\moduleRightDash\@xx{}%
\@pvspace{8.0pt}%
\@x{ {\EXTENDS} EgalitarianPaxos ,\, TLC}%
\@pvspace{8.0pt}%
\@x{}\midbar\@xx{}%
\@pvspace{8.0pt}%
\@x{ {\CONSTANTS} p1 ,\, p2 ,\, p3 ,\, c1 ,\, c2}%
\@x{ {\VARIABLES} HIndex}%
\@pvspace{8.0pt}%
\@x{ MCReplicas \.{\defeq}\@s{4.1} \{ p1 ,\, p2 ,\, p3 \}}%
\@x{ MCCommands \.{\defeq} \{ c1 ,\, c2 \}}%
 \@x{ MCFastQuorums ( X ) \.{\defeq} {\IF} X \.{=} p1 \.{\THEN} \{ \{ p1 ,\,
 p3 \} \}}%
\@x{ \.{\ELSE} {\IF} X \.{=} p2 \.{\THEN} \{ \{ p1 ,\, p2 \} \}}%
\@x{ \.{\ELSE} \{ \{ p2 ,\, p3 \} \}}%
\@x{ MCSlowQuorums ( X ) \.{\defeq} MCFastQuorums ( X )}%
\@x{ MCMaxBallot \.{\defeq} 5}%
\@pvspace{8.0pt}%
\@x{ AdvanceHistory ( pos ) \.{\defeq} \.{\land} HIndex \.{=} pos}%
\@x{ \.{\land} HIndex \.{'} \.{=} HIndex \.{+} 1}%
\@pvspace{8.0pt}%
\@x{ NewInit \.{\defeq}}%
\@x{\@s{8.2} \.{\land} HIndex \.{=} 1}%
\@x{\@s{8.2} \.{\land} sentMsg \.{=} \{ \}}%
\@x{\@s{8.2} \.{\land} cmdLog \.{=} [ r \.{\in} Replicas \.{\mapsto} \{ \} ]}%
\@x{\@s{8.2} \.{\land} proposed \.{=} \{ \}}%
 \@x{\@s{8.2} \.{\land} executed \.{=} [ r \.{\in} Replicas \.{\mapsto} \{ \}
 ]}%
\@x{\@s{8.2} \.{\land} crtInst \.{=} [ r \.{\in} Replicas \.{\mapsto} 1 ]}%
 \@x{\@s{8.2} \.{\land} leaderOfInst \.{=} [ r \.{\in} Replicas \.{\mapsto} \{
 \} ]}%
 \@x{\@s{8.2} \.{\land} committed \.{=} [ i \.{\in} Instances \.{\mapsto} \{
 \} ]}%
\@x{\@s{8.2} \.{\land} ballots \.{=} 1}%
 \@x{\@s{8.2} \.{\land} preparing \.{=} [ r \.{\in} Replicas \.{\mapsto} \{ \}
 ]}%
\@pvspace{8.0pt}%
 \@x{ NewNext \.{\defeq} \.{\lor} ( AdvanceHistory ( 1 ) \.{\land} Propose (
 c1 ,\, p3 ) )}%
\@x{ \.{\lor} ( AdvanceHistory ( 2 ) \.{\land} Propose ( c2 ,\, p1 ) )}%
\@x{ \.{\lor} ( AdvanceHistory ( 3 ) \.{\land} Phase1Reply ( p3 ) )}%
 \@x{ \.{\lor} ( AdvanceHistory ( 4 ) \.{\land} SendPrepare ( p3 ,\, {\langle}
 p1 ,\, 1 {\rangle} ,\, \{ p2 ,\, p3 \} ) )}%
\@x{ \.{\lor} ( AdvanceHistory ( 5 ) \.{\land} ReplyPrepare ( p2 ) )}%
\@x{ \.{\lor} ( AdvanceHistory ( 6 ) \.{\land} ReplyPrepare ( p3 ) )}%
 \@x{ \.{\lor} ( AdvanceHistory ( 7 ) \.{\land} PrepareFinalize ( p3 ,\,
 {\langle} p1 ,\, 1 {\rangle} ,\, \{ p2 ,\, p3 \} ) )}%
 \@x{ \.{\lor} ( AdvanceHistory ( 8 ) \.{\land} SendPrepare ( p2 ,\, {\langle}
 p1 ,\, 1 {\rangle} ,\, \{ p1 ,\, p2 \} ) )}%
\@x{ \.{\lor} ( AdvanceHistory ( 9 ) \.{\land} ReplyPrepare ( p1 ) )}%
\@x{ \.{\lor} ( AdvanceHistory ( 10 ) \.{\land} ReplyPrepare ( p2 ) )}%
 \@x{ \.{\lor} ( AdvanceHistory ( 11 ) \.{\land} PrepareFinalize ( p2 ,\,
 {\langle} p1 ,\, 1 {\rangle} ,\, \{ p1 ,\, p2 \} ) )}%
\@x{ \.{\lor} ( AdvanceHistory ( 12 ) \.{\land} Phase1Reply ( p1 ) )}%
 \@x{ \.{\lor} ( AdvanceHistory ( 13 ) \.{\land} Phase1Slow ( p2 ,\, {\langle}
 p1 ,\, 1 {\rangle} ,\, \{ p1 ,\, p2 \} ) )}%
\@x{ \.{\lor} ( AdvanceHistory ( 14 ) \.{\land} Phase2Reply ( p1 ) )}%
 \@x{ \.{\lor} ( AdvanceHistory ( 15 ) \.{\land} SendPrepare ( p1 ,\,
 {\langle} p1 ,\, 1 {\rangle} ,\, \{ p1 ,\, p3 \} ) )}%
\@x{ \.{\lor} ( AdvanceHistory ( 16 ) \.{\land} ReplyPrepare ( p3 ) )}%
 \@x{ \.{\lor} ( AdvanceHistory ( 17 ) \.{\land} SendPrepare ( p1 ,\,
 {\langle} p1 ,\, 1 {\rangle} ,\, \{ p1 ,\, p3 \} ) )}%
\@x{ \.{\lor} ( AdvanceHistory ( 18 ) \.{\land} ReplyPrepare ( p3 ) )}%
\@x{ \.{\lor} ( AdvanceHistory ( 19 ) \.{\land} ReplyPrepare ( p1 ) )}%
\@x{ \.{\lor} ( AdvanceHistory ( 20 ) \.{\land} ReplyPrepare ( p1 ) )}%
\@y{\@s{0}%
 answer both ballots
}%
\@xx{}%
 \@x{ \.{\lor} ( AdvanceHistory ( 21 ) \.{\land} PrepareFinalize ( p1 ,\,
 {\langle} p1 ,\, 1 {\rangle} ,\, \{ p1 ,\, p3 \} ) )}%
\@x{ \.{\lor} ( AdvanceHistory ( 22 ) \.{\land} Phase2Reply ( p3 ) )}%
 \@x{ \.{\lor} ( AdvanceHistory ( 23 ) \.{\land} Phase2Finalize ( p2 ,\,
 {\langle} p1 ,\, 1 {\rangle} ,\, \{ p1 ,\, p2 \} ) )}%
 \@x{ \.{\lor} ( AdvanceHistory ( 24 ) \.{\land} Phase2Finalize ( p1 ,\,
 {\langle} p1 ,\, 1 {\rangle} ,\, \{ p1 ,\, p3 \} ) )}%
\@pvspace{8.0pt}%
\@x{}\bottombar\@xx{}%

\subsection{Model}
\labappendix{counterexample:model}

\begin{lstlisting}
CONSTANTS
p1 = p1
p2 = p2
p3 = p3
c1 = c1
c2 = c2
none = none
Replicas <- MCReplicas
Commands <- MCCommands
FastQuorums <- MCFastQuorums
SlowQuorums <- MCSlowQuorums
MaxBallot <- MCMaxBallot
Init <- NewInit
Next <- NewNext
SPECIFICATION Spec
PROPERTY Consistency
\end{lstlisting}

\end{document}